\documentclass[fleqn,usenatbib]{mnras}
\usepackage{mathptmx}
\usepackage{float}

\usepackage[T1]{fontenc}
\usepackage{ae,aecompl}

\usepackage{graphicx}	% Including figure files
\usepackage{amsmath}	% Advanced maths commands
\usepackage{amssymb}	% Extra maths symbols
\usepackage[normalem]{ulem}
\usepackage{soul}
\usepackage{multirow}
\usepackage{mathtools}
\usepackage{physics}
\usepackage{lmodern}
\usepackage{xspace} % this decides if space is needed or not

\newcommand{\AG}{{\small AX-GADGET}\xspace}

\newcommand{\SUBFIND}{{\small SUBFIND}\xspace}
\newcommand{\AC}{{\small axionCAMB}\xspace}

\newcommand{\main}{halo\xspace}
\newcommand{\sats}{satellites\xspace}

\newcommand{\mratio}{\mathrm{\tilde{m}}}

\usepackage{mathtools}
\usepackage{siunitx}
\AtBeginDocument{\RenewCommandCopy\qty\SI}

\usepackage{amsmath}
\usepackage [english]{babel}
\usepackage [autostyle, english = american]{csquotes}
\MakeOuterQuote{"}
\title[No Catch-22 for FDM]{No \emph{Catch-22} for Fuzzy Dark Matter: testing substructure counts and core sizes via high-resolution cosmological simulations}

\author[S. Elgamal et al.]{Sana Elgamal$^{1,2}$\thanks{E-mail: selgamal@uchicago.edu}, Matteo Nori$^{1,2}$, Andrea V. Macciò$^{1,2,3}$, Marco Baldi$^{4,5,6}$,\newauthor{Stefan Waterval}$^{1,2}$
\\
$^{1}$New York University Abu Dhabi, PO Box 129188 Saadiyat Island, Abu Dhabi, United Arab Emirates\\
$^{2}$Center for Astrophysics and Space Science (CASS), New York University
Abu Dhabi, PO Box 129188, Abu Dhabi, UAE\\
$^{3}$Max Planck Institut f\"{u}r Astronomie, K\"{o}nigstuhl 17, D-69117 Heidelberg, Germany\\
$^{4}$Dipartimento di Fisica e Astronomia, Alma Mater Studiorum - University of Bologna, Via Piero Gobetti 93/2, 40129 Bologna BO, Italy\\
$^{5}$INAF - Osservatorio Astronomico di Bologna, Via Piero Gobetti 93/3, 40129 Bologna BO, Italy\\
$^{6}$INFN - Istituto Nazionale di Fisica Nucleare, Sezione di Bologna, Viale Berti Pichat 6/2, 40127 Bologna BO, Italy\\
}

\date{Accepted XXX. Received YYY; in original form ZZZ}

\pubyear{2023}
\begin{document}
\label{firstpage}
\pagerange{\pageref{firstpage}--\pageref{lastpage}}
\maketitle

% Abstract of the paper
\begin{abstract}
Fuzzy Dark Matter (FDM) has recently emerged as an interesting alternative model to the standard Cold Dark Matter (CDM). In this model, dark matter consists of very light bosonic particles with quantum mechanical effects on galactic scales. 
Using the \textit{N}-body code \AG, we perform cosmological simulations of FDM that fully model the dynamical effects of the quantum potential throughout cosmic evolution. Through the combined analysis of FDM volume and high-resolution zoom-in simulations of different FDM particle masses  ($m_{\chi}$ $\sim$ $10^{-23} - 10^{-21}$ eV/c$^2$), we study how FDM impacts the abundance of substructure and the inner density profiles of dark matter haloes. For the first time, using our FDM volume simulations, we provide a fitting formula for the FDM-to-CDM subhalo abundance ratio as a function of the FDM mass. 

More importantly, our simulations clearly demonstrate that there exists an extended FDM particle mass interval able to reproduce the observed substructure counts and, at the same time, create substantial cores ($r_{c} \sim 1$ kpc) in the density profile of dwarf galaxies ($\approx 10^{9}-10^{10}$ M$_{\sun}$), which stands in stark contrast with CDM predictions even with baryonic effects taken into account. The dark matter distribution in the faintest galaxies offers then a clear way to discriminate between FDM and CDM. 
\end{abstract}


\section{Introduction}
\label{sec:intro}
%the case for dark matter
Evidence for dark matter has accumulated from an array of observations spanning a wide range of scales. On galactic scales, the observed rotation curves of galaxies remain relatively flat at large radii, which significantly deviates from the theoretical expectation derived solely from the visible matter \citep{1980ApJ...238..471R, bosma198121}. On cluster scales, the mass-to-light ratios of galaxy clusters are significantly larger than unity \citep{1937ApJ....86..217Z, bahcall1995dark}, providing evidence for the presence of a massive non-luminous component. Moving on to cosmological scales, observations of the large-scale structure \citep{PhysRevD.74.123507} require the existence of a non-relativistic clustering component that decoupled from the hot primordial plasma well before recombination in order to explain the highly non-linear structure that we observe in the Universe today. In addition, the observed Cosmic Microwave Background (CMB) anisotropies \citep{2011ApJS..192...18K, 2016A&A...594A..13P} along with observations from Type IA Supernovae \citep{1998AJ....116.1009R, 1999ApJ...517..565P} constrain the total matter energy density in the Universe, which has been found to greatly exceed the baryon density from Big Bang Nucleosynthesis \citep{copi1995big}. This provides further evidence for the existence of an additional gravitational component of non-baryonic origin. Taken together, these independent observations have built a compelling case for the presence of a non-luminous, non-baryonic, clustering component that dominates the matter content of the Universe.

The current picture of cosmological structure formation heavily relies on the existence of the so-called Cold Dark Matter (CDM), whose physical nature remains unknown. While the standard hierarchical CDM picture has proven to provide a remarkably successful description of large-scale cosmological observations \citep{PhysRevD.74.123507, aghanim2020planck, 2017MNRAS.470.2617A}, assessing its validity on highly nonlinear scales has proven to be a much more challenging task. On the theoretical front, baryonic physics is known to play a significant role on those scales \citep{brooks2014baryons, onorbe2015forged,Maccio2020,Waterval2022}, with effects that are difficult to disentangle from those arising from the fundamental nature of dark matter. On the observational side, the inefficiency of star formation particularly in dwarf galaxies \citep{2003MNRAS.339.1057Y, 2017MNRAS.471.3547F, Frings2017} renders their detection difficult, making it harder to perform a robust measurement of the subhalo mass function of the Local Group. This, combined with the lack of direct detection of the dark matter particles in the $\mathcal{O}$(GeV/c$^2$) mass scale has motivated efforts to formulate and explore models beyond the standard CDM paradigm. Since the success of CDM is extremely well-established on linear scales, it follows that any alternative model of dark matter needs to match the behaviour of CDM on large scales in order to constitute a viable model of structure formation.

One such particle dark matter alternative is Fuzzy Dark Matter (FDM), which~--~with motivations from particle theory~--~posits that dark matter consists of extremely light bosons \citep{Hu00}. The bosons possess a very small mass which \citet{Hu00} originally proposed to be $\mathcal{O}(10^{-22}$ eV/c$^2$) \citep[see also][for the first limits on FDM]{Amendola2006}. This extremely low mass corresponds to a de Broglie wavelength on scales of $\mathcal{O}(1$  kpc), meaning that the wave-like behaviour of FDM manifests on those sub-galactic scales \citep[see e.g.][for comprehensive reviews]{PhysRevD.95.043541, ferreira2021ultra}.

The quantum nature of FDM is encapsulated in a potential form referred to as "quantum potential" which arises due to the uncertainty principle. This quantum potential counteracts gravity and hence stabilizes the small-scale density perturbations against collapse. Therefore, unlike CDM, where the density perturbations are unstable and thus able to collapse on all spatial scales below the Hubble horizon, the quantum potential in FDM introduces a characteristic scale below which perturbations are stable. This gives rise to a suppression of the formation of low-mass haloes and subhaloes, which effectively translates to a sharp reduction in the FDM linear power on small scales. In addition to this, the quantum potential also impacts the internal structure of FDM subhaloes. In fact, FDM subhaloes naturally have cored (i.e., flat) instead of cuspy (i.e., with density diverging as $\propto 1/r$) inner density profiles, since the quantum potential limits the central density of collapsed subhaloes \citep{PhysRevD.95.043541}.

%FDM particle mass constraints

The mass of the FDM particle --~denoted hereafter as $m_{\chi}$~-- is the only free parameter in the FDM model. To ensure that the field behaves like dark matter while remaining relevant in cosmological structure formation and reionization, the mass of the FDM particle should fall within the range $10^{-33}$ eV/c$^2$ $\lesssim m_{\chi} \lesssim 10^{-18}$ eV/c$^2$. On linear scales, observational signatures constrain the FDM particle mass to $m_{\chi} \gtrsim 10^{-24}$ eV/c$^2$ \citep{MARSH20161}. Since the quantum potential term only dominates below scales comparable to the de Broglie scale, deriving useful constraints on $m_{\chi}$ requires investigation of the non-linear scales. While those scales in principle possess rich cosmological information that can be utilised to further constrain the FDM mass, they are nonetheless complicated by effects arising from baryonic physics which are still only partially understood. Despite this uncertainty, efforts are under way to establish constraints on $m_{\chi}$ from the non-linear scales. \citet{amorisco2018first} used stellar streams to constrain the FDM particle mass to be $m_{22} \gtrsim 1.5$ (hereafter $m_{22} \equiv m_{\chi}/{10^{-22} \text{eV}/\text{c}^2}$). Furthermore, since the quantum potential delays the onset of structure formation and hence galaxy formation in FDM, \citet{kulkarni2022halo} utilized observations of high-redshift lensed galaxies to impose a constraint of $m_{22} \gtrsim 2$ on the FDM mass. 

 Additionally, observations of the Lyman-$\alpha$ forest trace the dark matter distribution in the high-redshift Universe, allowing the inference of the matter power spectrum on scales 0.3 $h$ Mpc$^{-1}$ $\lesssim$ k $\lesssim$ 12.5 $h$ Mpc$^{-1}$ and thus providing a robust means of testing and constraining the nature of dark matter. Accounting for the full non-linear treatment of the quantum potential, measurements of the 1D Lyman-$\alpha$ forest flux power spectrum yielded a lower bound of $\approx 21.08 \times 10^{-22}$ eV/c$^2$ on the FDM mass at the 2$\sigma$ level \citep{Nori19}.

In a variant to CDM (that is often considered to be akin to FDM) known as Warm Dark Matter (WDM), dark matter consists of standard collisionless particles whose mass is sufficiently small $\sim \mathcal{O}(1$ keV/c$^2$) that its primordial velocity dispersion significantly impacts structure formation through free-streaming, suppressing the formation of substructure and inducing core formation in dwarf galaxies \citep{2005PhRvD..71f3534V, abazajian2006production, lovell2014properties}. 

However, \cite{Maccio12} demonstrated the presence of a \textit{Catch-22} problem within the framework of a WDM universe. This refers to the absence of a WDM particle mass interval that can account for the observed abundance of substructures and the substantial density cores inside dwarf galaxies \textit{simultaneously}. In fact, in order to form the observed significant core ($\sim 1$ kpc) inside a dwarf galaxy, a thermal candidate mass of $\sim$ 0.1 keV/c$^2$ is required. This mass already prevents the formation of the galaxy through free-streaming. On the other hand, to form the dwarf galaxy itself, a candidate mass range of $\sim 1-2$ keV/c$^2$ is needed, but this range fails to produce the observed large cores. 

Using numerical simulations of structure formation that incorporate the characteristic non-linear interaction of FDM dynamics, and by computing the substructure abundance and inner density profiles of subhaloes for different FDM particle masses, the goal of this work is to test whether a similar \textit{Catch-22} problem exists in an FDM universe.

\label{sec:fdm_mass}
The rest of this paper is structured as follows. In Sec.~\ref{sec:theory} we review the theoretical background of FDM cosmology. In Sec.~\ref{sec:NM} we detail our numerical methodology. In Sec.~\ref{sec:results} we present our FDM Halo Mass Function and SubHalo Mass Function (hereafter HMF and SHMF, respectively) as measured from our cosmological volume simulations. We also present the FDM radial density profiles which are derived from our zoom-in simulations. We then use observations of the SHMF and density profiles of dwarf galaxies surrounding the Milky Way in conjunction with our results to assess whether there is a \textit{Catch-22} problem in FDM. A summary of our findings and conclusions is provided in Sec.~\ref{sec:conclusions}. Additionally, we present the FDM concentration-mass relation computed using our volume simulations in Appendix~\ref{subsec: conc_mass}.

\section{Theory}
\label{sec:theory}
 The action of the scalar field $\phi$ describing FDM takes the following form:
 \begin{equation}
\label{eq:action}
 S = \int \frac{d^{4}x}{\hslash c^{2}}{\sqrt{-g}} \left \{\frac{1}{2} g^{\mu \nu} \grad_{\mu} \phi \grad_{\nu} \phi\ - \frac{1}{2} \frac{m_{\chi}^{2}c^{2} }{ {\hslash}^{2}}{\phi}^{2} \right \},
\end{equation}
where $g^{\mu \nu}$ is the metric tensor for the perturbed Friedmann-Lema\^{i}tre-Robertson-Walker (FLRW) metric given by:
\begin{equation}
\label{eq:FLRW}
ds^{2} = \left(1 + \dfrac{2\Phi}{c^2} \right) c^2dt^{2} - a^{2}(t) \left(1 - \dfrac{2 \Phi}{c^2} \right) d\mathbf{r}^{2},
\end{equation}
where $\Phi(\mathbf{r}, t)$ denotes the Newtonian gravitational potential and $a(t)$ is the scale factor. In the non-relativistic limit (which is relevant in the context of structure formation), it is convenient to express the real field $\phi$ in terms of another complex scalar field $\psi$:
\begin{equation}
\label{eq:psi}
\phi = \sqrt{\frac{\hslash^3 c}{2m_{\chi}}} \left ( \psi e^{-im_{\chi} c^2 t/\hslash} + \psi^{\ast} e^{im_{\chi}c^2 t/\hslash} \right).
\end{equation}

Using the non-relativistic assumption that $\psi$ is slowly changing with time $|\ddot{\psi}| << \dfrac{m_{\chi}}{\hslash}c^2|\dot{\psi}|$ (where dots denote cosmic time derivatives), one obtains  the following equation of motion:
\begin{equation}
\label{eq:SP}
i {\hslash} \left (\dot{\psi} + \frac{3}{2} H \psi \right) =  \left ( - \frac{ \hslash^2}{2m_{\chi}a^2} \nabla^2 + m_{\chi} \Phi \right) \psi,
\end{equation}
which takes the form of the Schr\"{o}dinger equation in an expanding universe. Here $H(t) \equiv \frac{\dot{a}}{a}$ is the Hubble parameter.

When studying cosmological structure formation, it is often useful to adopt the fluid approach. By performing a Madelung transformation \citep[][]{Madelung27}, the equations can be reformulated from a field description to a fluid description. We define the dark matter fluid density $\rho$ and velocity $\mathbf{v}$ by:
\begin{equation}
\label{eq:psi_madelung}
\rho \equiv m|\psi|^2, \hspace{0.3cm} \mathbf{v} \equiv \frac{\hslash}{am} \grad \theta = \frac{-i\hslash}{2ma} \left (\frac{\grad \psi}{\psi} -  \frac{\grad \psi^{\ast}}{\psi^{\ast}} \right),
\end{equation}
 where $\theta$ represents the phase in $\psi = \sqrt{\frac{\rho}{m}}e^{i \theta}$. One can then recast the Schr\"{o}dinger equation (Eq.~\ref{eq:SP}) into the following fluid equations (Eqs.~\ref{eq:continuity} and \ref{eq:euler}) in comoving coordinates:
\begin{equation}
\label{eq:continuity}
\dot{\rho} + 3H \rho + \frac{1}{a} \grad \cdot (\rho \mathbf{v}) = 0,
\end{equation}
\begin{equation}
\label{eq:euler}
\dot{\mathbf{v}} + H \mathbf{v} + \frac{1}{a} (\mathbf{v} \cdot \grad) \mathbf{v} = - \frac{1}{a} \grad{\Phi} + \frac{\hslash^2}{2a^3m_{\chi}^2} \grad(\frac{\nabla^2 \sqrt{\rho}}{\sqrt{\rho}}),
\end{equation}
by equating the imaginary and real parts of Eq.~\ref{eq:SP}, respectively. Eq. ~\ref{eq:continuity} is the continuity equation while Eq.~\ref{eq:euler} resembles an Euler equation with an additional quantum potential $Q$ term defined by:
\begin{equation}
\label{eq:quantum pressure}
Q \equiv \frac{\hslash^2}{2m_{\chi}^2} \left ( \frac{\nabla^2 \sqrt{\rho}}{\sqrt{\rho}} \right) = \frac{\hslash^2}{4m_{\chi}^2} \left( \frac{\nabla^2 \rho}{\rho} - \frac{|\grad {\rho}|^2}{2 \rho^{2}} \right).
\end{equation}

To describe the behavior of the gravitational potential in response to fluctuations in the dark matter density field, Eqs. ~\ref{eq:continuity} and ~\ref{eq:euler} need to be coupled to the Poisson equation:
\begin{equation}
\label{eq:poisson}
\nabla^2 \Phi = 4 \pi G a^2 \Bar{\rho} \delta ,
\end{equation}
where $\delta \equiv \frac{\rho - \bar{\rho}}{\bar{\rho}}$ denotes the density contrast relative to the background density $\Bar{\rho}$. Together with Eqs.~\ref{eq:continuity} and~\ref{eq:euler}, they comprise the Schr\"{o}dinger-Poisson system. Perturbing Eqs. ~\ref{eq:continuity} and ~\ref{eq:euler} around $\rho = \bar{\rho}$, $|\mathbf{v}|=0$, and $\Phi = 0$, the density contrast in Fourier space~--~denoted as $\delta_{k}$~--~satisfies the following equation (in the comoving frame) for a harmonic oscillator with a time-dependent mass term due to both gravity and the quantum potential as well as a friction term due to the Hubble expansion:
\begin{equation}
\label{eq:SHO}
\ddot{\delta}_{k} + 2H \dot{\delta}_{k} + \left (\frac{\hslash^2 k^4}{4m_{\chi}^2 a^4} - \frac{4 \pi G \rho_{0}}{a^3} \right) \delta_{k} = 0,
\end{equation}
where $k$ is the comoving wavenumber and $\rho_{0} = \bar{\rho} a^{3}$ is the comoving FDM background density. From the negative sign in the mass term in Eq.~\ref{eq:SHO}, one observes that in addition to the usual gravitational term which reinforces the growth of FDM perturbations, the extra term due to the quantum potential counteracts such growth. Evidently, the FDM perturbations grow if gravity wins over the quantum potential $\left( \frac{\hslash^2 k^4}{4m_{\chi}^2 a^4} < \frac{4 \pi G \rho_{0}}{a^3} \right)$. One can then readily identify the characteristic scale --~referred to as the quantum Jeans scale --~at which the gravitational attraction exactly balances the repulsion due to the quantum potential:
\begin{equation}
\label{eq:Jeans_scale}
k_{J} = \left (\frac{16 \pi G \rho_{0} m_{\chi}^{2}}{\hslash^2} \right)^{1/4} a^{1/4}.
\end{equation}
 This means that the quantum potential in FDM introduces a threshold radius below which perturbations are unable to collapse. A corresponding quantum Jeans mass is then defined as the mass enclosed within the comoving radius $r = \frac{\pi}{k_{J}}$:
\begin{equation}
\label{eq:Jeans_mass}
M_{J} = \frac{4 \pi}{3} \rho_{0} \left (\frac{\pi}{k_{J}} \right)^{3} \propto a^{-3/4} m_{\chi}^{-3/2},
\end{equation}
 and it represents the characteristic mass scale below which gravitational collapse and hence structure formation is prevented by FDM interaction at a particular redshift.

To find the growing solution for Eq.~\ref{eq:SHO}, we factorize  $\delta_{k}$ into a factor that depends only on $k$ and another time-dependent factor $D(t)$ by writing $\delta_{k} = A_{k} D(t)$. Plugging this in Eq.~\ref{eq:SHO} gives the growing mode solution:
\begin{equation}
\label{eq:growing_mode}
  D_{+}(t) \hspace{0.1cm} \propto 
  \begin{cases}
    a, \hspace{0.1cm} k \ll k_{J} \\
    1, \hspace{0.1cm} k \gg k_{J}\\
  \end{cases}.
\end{equation}

From Eq.~\ref{eq:growing_mode}, it can be inferred that structure formation in FDM is identical to that in CDM (where $D_{+}(t) \propto a$ for all $k$ modes) on large scales ($k \ll k_{J}$), but~--~as already discussed~--~is suppressed on small scales ($k \gg k_{J}$). This implies that the linear power spectrum of FDM is expected to match CDM on large scales. Indeed, it has been demonstrated using numerical simulations of structure formation that the large-scale structure of FDM is indistinguishable from that of CDM \citep{schive2014cosmic}.
\label{sec:fdm_th}

In contrast to linear perturbation theory, where the growth of small-scale FDM density perturbations is evidently unable to catch up with that of the large-scale perturbations, non-linear theory predicts an enhanced growth of those small-scale perturbations. This enhancement arises from the interference between different $k$ modes, which becomes particularly evident in the non-linear regime. As a consequence, density perturbations of order unity emerge on the de Broglie scale \citep{hui2021wave}. Indeed, cosmological simulations of FDM have demonstrated that perturbations on the quasi-linear and non-linear scales eventually catch up with the large-scale ones (at low redshifts), leading to a reduction in the amount of power suppression relative to CDM on those scales \citep{Marsh16nl,Nori18}.

\subsection{Density cores}

The Schr\"{o}dinger-Poisson system (when solved along with the appropriate boundary conditions) can be shown to have a stable non-degenerate solution~--~i.e., the core~--~with different energy densities. The ground state solution is referred to as the "soliton" \citep{Hui16, Chavanis11a} and is characterized by the core density $\rho_{c}$ and core radius $r_{c}$, which are defined as the finite density at $r = 0$ and the radius at which the density drops by one-half its value at $r = 0$ (i.e., $\rho(r_{c}) \equiv \rho_{c}/2$), respectively.

This soliton solution is found in the innermost regions of FDM subhaloes, which exhibit a flat central core. This is in contrast to the standard CDM subhaloes, whose density profiles are reasonably well-fitted by the Navarro-Frenk-White (\cite{Navarro1996}, hereafter NFW) profile given by:
\begin{equation}
\label{eq:NFW}
 \rho(r) = \rho_{s}\,\left(\frac{4\,r_{s}^{3}}{r\,(r + r_{s})^2} \right),
\end{equation}
where $r_{s}$ and $\rho_{s}$ denote the characteristic scale radius and density at which the logarithmic density slope is $-2$, respectively. Evidently, in the limit where $r \ll r_{s}$, the logarithmic density slope approaches $-1$, while it approaches $-3$ for $r \gg r_{s}$.

 Since the FDM quantum potential ceases to dominate over distances larger than the quantum Jeans scale, the FDM subhaloes typically exhibit an NFW envelope as in CDM. Consequently, FDM subhaloes are accurately described by a cored-NFW density profile \citep{chan2022diversity, schive2014understanding}, which is given by:
\begin{equation}
\label{eq:cored-NFW}
\rho(r)=
\begin{cases}
    \rho_{c} \left[ 1 + \left( \sqrt[8]{2} - 1 \right) \left( \dfrac{r}{r_c} \right)^2 \right]^{-8} & { \text{$r<r_{t}$}}\\[10pt]
    4 \rho_s \left( \dfrac{r}{r_s} \right)^{-1} \left(1 + \dfrac{r}{r_s} \right)^{-2} & \text{$r \ge r_{t}$}
\end{cases},
\end{equation}
where $r_{t}$ denotes the radius at which the cored density profile transitions to an NFW.

The Schr\"{o}dinger-Poisson system has an important scaling symmetry, where the system is invariant under the transformation $\{\tau, x, \psi, \Phi\} \rightarrow \{\lambda^{-2} \tau, \lambda^{-1} x, \lambda^{2} \psi, \lambda^{2} \Phi\}$ for an arbitrary scale factor $\lambda$ \citep{Ji94}. This corresponds to the following transformation of the relevant physical quantities $\{r_{c}, \rho_{c}\} \rightarrow \{\lambda^{-1} r_{c}, \lambda^{4} \rho_{c}\}$, which then fixes the core-scaling relation between $r_{c}$ and $ \rho_{c}$ as follows:
\begin{equation}
\label{eq:rcmrhoc_scaling}
 r_{c} \propto (m_{\chi}^{2}\,\rho_{c})^{-1/4}.
\end{equation}

\section{Numerical Methods}
\label{sec:NM}

To simulate structure and substructure formation, we use the \textit{N}-body code \AG which is based on a particle-oriented solution of the Schr\"{o}dinger-Poisson system and incorporates the quantum effects of FDM throughout cosmic evolution \citep[see][for a detailed description of the code]{Nori18}.

To study the structure and substructure of FDM, we employ two distinct types of simulations. We utilize a set of eight cosmological volume simulation boxes, consisting of one CDM and seven FDM boxes with particle masses $m_{22} \in \{0.25, 0.50, 1.00, 2.00, 4.00, 8.00, 16.0\}$\footnote{Note that, to preserve the symmetry of \AG internal units and their dependence on $h = H_0 / 100$ km s$^{-1}$ Mpc$^{-1}$, the conversion to the physical FDM particle mass is $m_{\chi} = h^{-1/2} \ m_{22} \ 10^{-22} \text{eV}/\text{c}^2$.}. In addition, we use a set of sixteen zoom-in simulations, representing four different low-mass systems, and each run using three FDM particle masses $m_{22} \in \{8.00, 16.0, 32.0\}$ along with their CDM counterpart. The FDM component comprises the entirety of the dark matter content within our simulations.

Those two different simulations sets together enable us to simultaneously attain sufficient large-scale resolution to obtain accurate HMF and SHMF averaged over a statistically large sample, as well as high small-scale resolution to characterize the features of the innermost density cores. Our choice of the FDM mass range $m_{22} \in [0.25 - 32.0]$ is motivated by its relevance at the scales of interest to both the SHMF and the density profiles.

Our volume and zoom-in simulation sets adopt cosmological parameters consistent with \cite{2016A&A...594A..13P}: matter density $\Omega_{m} = 0.317$, dark energy density $\Omega_{\Lambda} = 0.683$, Hubble constant $H_{0} = 67.3$ km s$^{-1}$ Mpc$^{-1}$, primordial matter power spectrum slope $n_{s} = 0.96$ and linear matter power spectrum normalization $\sigma_{8} = 0.829$. 

In order to ensure that the volume simulations represent the same volume of the Universe, all of the volume simulations in this work share the same seed for the random generation of the density perturbations. The specific distribution of perturbations is picked in accordance to the density power spectrum, which differs for CDM and each of the FDM models. We use the code \AC \citep{axionCAMB} to compute the power spectra in the FDM scenario. The volume simulation boxes are cubical in shape with a comoving side length of $20$~$h^{-1}$ Mpc, 
 featuring $512^3$ particles with a softening length of $300$~$h^{-1}$~pc and a particle mass of $5.24 \cdot 10^6$~$h^{-1}$M$_{\sun}$.

\subsection{Halo identification, merger tree construction and fragmentation correction}
\label{sec:subfind}

The identification of collapsed objects and construction of the merger tree were performed employing the same methods as in \citet{Aquarius}, based on the \SUBFIND code. The \SUBFIND code uses a Friends-of-friends algorithm \citep[][]{Davis_etal_1985} to identify halo systems, which are then subdivided and disentangled into subhaloes based on binding energy considerations. In our analysis, we only include subhaloes that have a minimum of $128$ particles. In the remainder of this text, we will refer to the overall dark matter system centered in the potential minimum as the \main, while the substructures within the \main are referred to as the \sats.

It is well-known that \textit{N}-body simulations with suppressed initial power spectrum at small scales suffer from the so-called "numerical fragmentation" problem \citep[]{Wang_White_2007}. This refers to the formation of small collapsed objects due to numerical artifacts rather than the gravitational evolution of primordial overdensities.

To mitigate such contamination from spurious subhaloes, we utilize the same strategy used in \citet{Nori19} \citep[based in turn on][]{Wang_White_2007,Lovell13} which imposes cutoffs on the number of particles, mass and shape of FDM subhaloes. The remaining sample can then be reliably attributed to the evolution of overdensities.

The cutoff for the number of particles is set to $1000$. For the mass cutoff scale estimate~--~denoted as $M_{\text{CUT}}$~--~we take into account the wavenumber at which the dimensionless power spectrum exhibits the highest amplitude $k_{\text{peak}}$ along with the inter-particle distance $d$ \citep{Lovell13,Wang_White_2007}:
\begin{equation}
\label{eq:MLIM}
M_{\text{CUT}} \sim 5\ \bar{\rho}\ \dfrac{ d}{k^2_{\text{peak}}}.
\end{equation}
Since we have FDM simulations with different particle masses (and hence different $k_{\text{peak}}$ values), this scale cut-off will accordingly vary depending on the masses. Above this mass scale, the majority of the haloes can be safely considered to have a physical origin. %In the case of our FDM simulation, this value is approximately $M_{CUT} \sim 5 \times 10^8 \dimM$ representing $\sim 10^4$ particles (a factor $10$ higher than the cut-ff in the number of particles suggested by \citet{Lovell13}).
As for the shape cut-off, we trace the halo particles back to their original positions in the initial conditions and use the sphericity $s$ of their initial spatial distribution (i.e. the ratio of the minor and major semi-axes) as our constraint, which is obtained from the inertia tensor of the equivalent triaxial ellipsoid. We apply the sphericity threshold value $s_{\text{CUT}}=0.16$ as determined by \citet{Lovell13} in the initial condition, below which the haloes are considered spurious. This value has also been verified to be valid in the FDM framework \citep{Nori19}. Table~\ref{tab:nhaloes_numfrag} reports the $M_{\text{CUT}}$ values used, as well as the total number of haloes (i.e., including those originating from numerical fragmentation) and genuine haloes in the FDM simulations with different masses. For general reference~--~even though CDM is not affected by numerical fragmentation~--~, we report that the number of haloes satisfying the number of particles and  sphericity cutoff in CDM is $4205$.

\begin{table}
\centering
\begin{tabular}{ c  c  c  c }
\hline
Model & $N_{\text{haloes}}$ & $N_{\text{haloes}}$ & $M_{\text{CUT}}$ \\
& & genuine & [$10^{10}$ $h^{-1}$ M$_{\sun}$] \\
\hline
FDM-0.25 & 1803 & 226 & 1.00 \\
FDM-0.50 & 2123 & 481 & 0.550 \\
FDM-1.00 & 2375 & 903 & 0.300 \\
FDM-2.00 & 3243 & 1635 & 0.165 \\
FDM-4.00 & 5481 & 2474 & 0.0886 \\
FDM-8.00 & 9727 & 3077 & 0.0481 \\
FDM-16.0 & 15047 & 3309 & 0.0262 \\
FDM-32.0 & 19310 & 3351 & 0.0142 \\
\hline
\end{tabular}
\caption{\label{tab:widgets2} The total number of haloes, genuine haloes, and the mass cutoff for all the FDM simulations with different masses.}
\label{tab:nhaloes_numfrag}
\end{table}

\section{Results}
\label{sec:results}

\begin{figure*}
\includegraphics[width=\textwidth, trim={0cm 0cm 0cm 0cm}, clip] {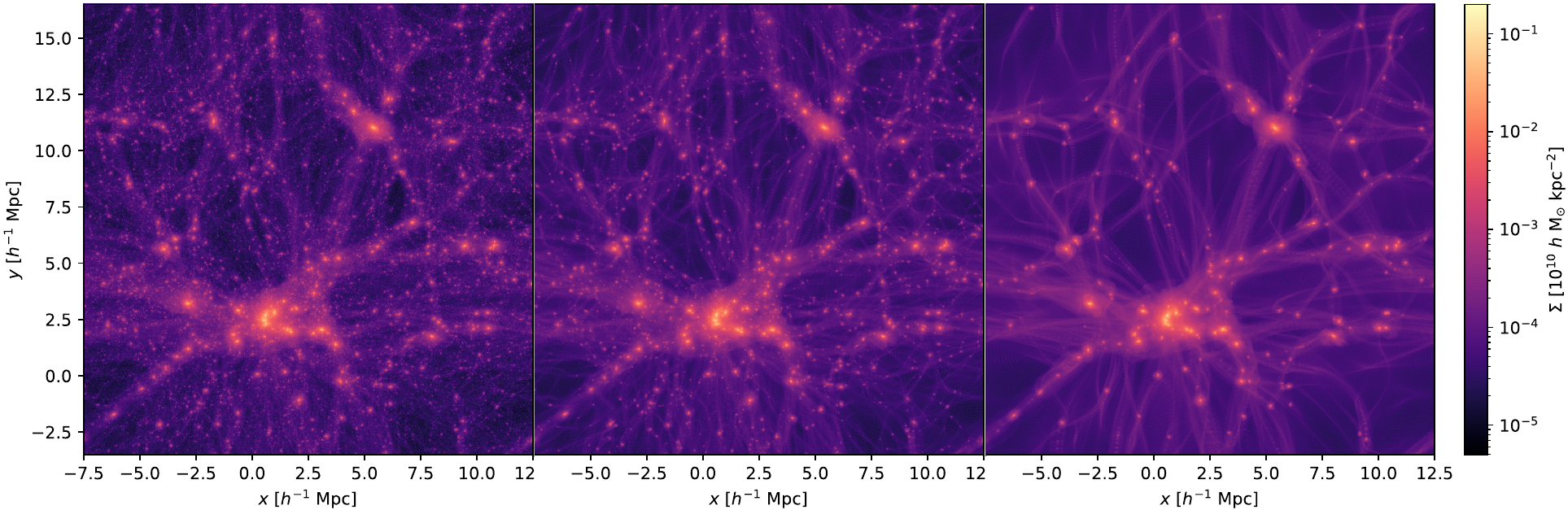}
\caption{Dark matter density maps of the volume simulations. From left to right: CDM, FDM with $m_{22}=2$ and FDM with $m_{22}=0.25$.}
\label{fig:maps}
\end{figure*}

\subsection{Halo Mass Function}

\begin{figure*}
\includegraphics[width=400pt, trim={0.16cm 0.11cm 0.11cm 0.11cm}, clip]{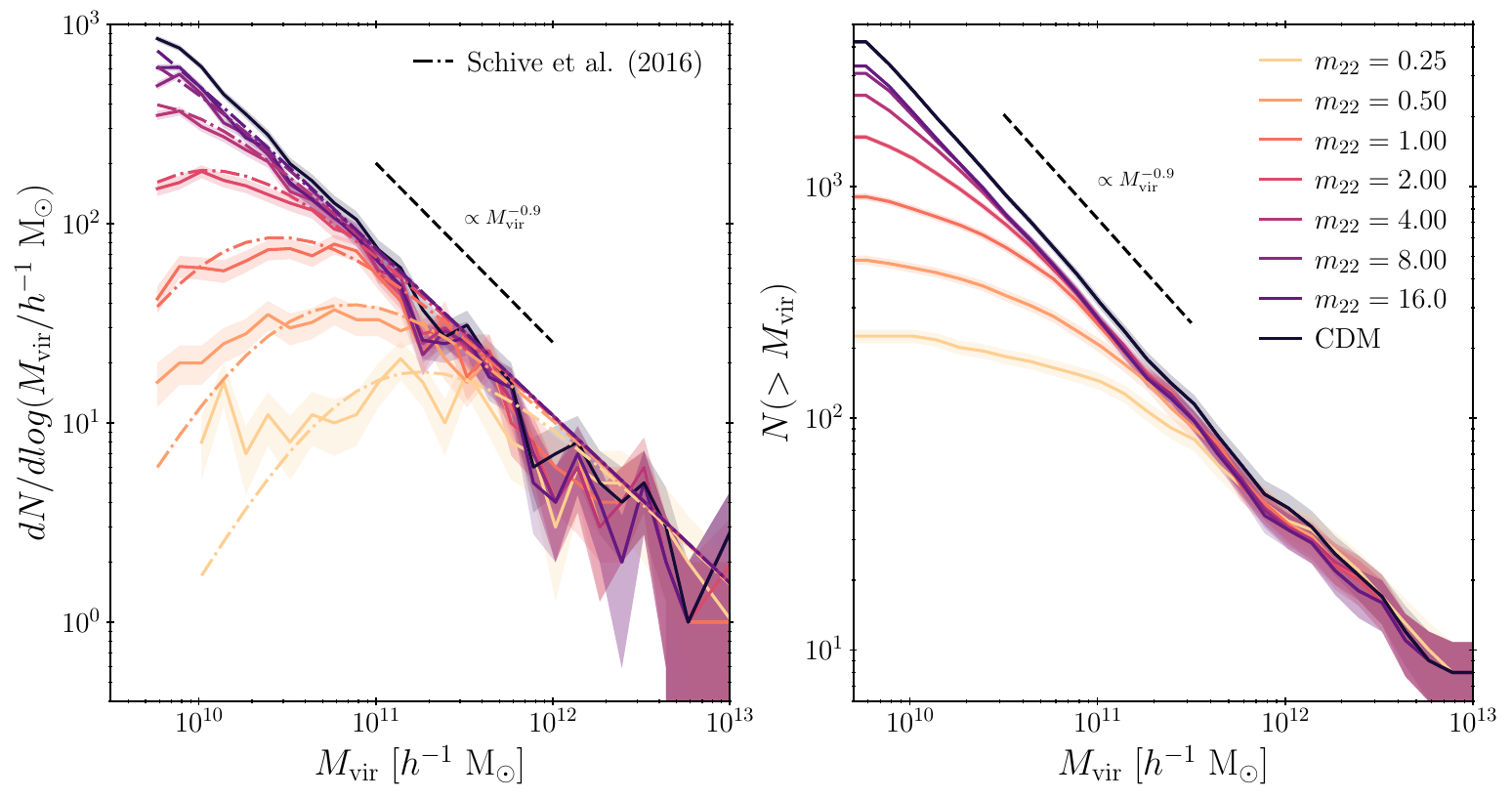}
\caption{The CDM and FDM differential (left) and cumulative (right) HMF for different FDM masses. The dashed black line is the HMF scaling obtained in \citet{springel2008aquarius} for CDM. The HMF given by \citet{schive2016contrasting} fitting formula are shown by the dash-dotted curves on the left plot. The $1 \sigma$ statistical errors are shown as shaded regions.}
\label{fig:hmf}
\end{figure*}

The change of the abundance of haloes across different dark matter models is evident in Fig.~\ref{fig:maps}, where dark matter density maps are shown for the CDM, the $m_{22}=2$ and the $m_{22}=0.25$ models, respectively.
We quantify the abundance of FDM (and CDM) haloes as a function of their virial mass as measured from our cosmological volume simulations. Fig.~\ref{fig:hmf} shows the differential (left) and cumulative (right) HMF derived from our FDM volume simulations at $z=0$ for seven FDM particle masses. We verified that the measured CDM HMF matches the theoretical scaling of $\propto M_{\text{vir}}^{-0.9}$ \citep{springel2008aquarius}, shown by the dashed black line. The HMF computed using the fitting formula given by \cite{schive2016contrasting} is plotted in dash-dotted curves for each FDM particle mass. As can be seen from Fig.~\ref{fig:hmf}, the FDM HMFs match the CDM one on large scales, with the cutoff scale varying depending on the FDM mass. As expected, the suppression of structure is stronger the lower the particle mass (as $Q \propto 1/m_{\chi}^2$).

\subsection{Subhalo Mass Function}

We quantify the abundance of subhaloes using the cumulative SHMF, which represents the number of subhaloes that are found in a system above a given mass. To make full use of the large number of systems present in each volume simulation, we average the SHMFs across all systems falling within three specific virial mass bins. One bin~--~hereafter referred to as the MW-like bin~--~represents the range that includes the Milky-Way halo virial mass $10^{11.5}$ M$_{\sun} \leq M_\text{vir} \leq 10^{12.5}$ M$_{\sun}$. The other two mass bins~--~hereafter referred to as bin $3$ and bin $1$~--~encompass the ranges above and below, respectively. For consistency in the intra-bin average among systems of different masses, we express the SHMF in terms of the satellite-to-virial mass ratio $\mratio \equiv m_\text{sat} / M_\text{vir}$ before averaging over the different systems.

\begin{figure*}
\includegraphics[width=\textwidth, trim={0.16cm 0.11cm 0.11cm 0.11cm}, clip] {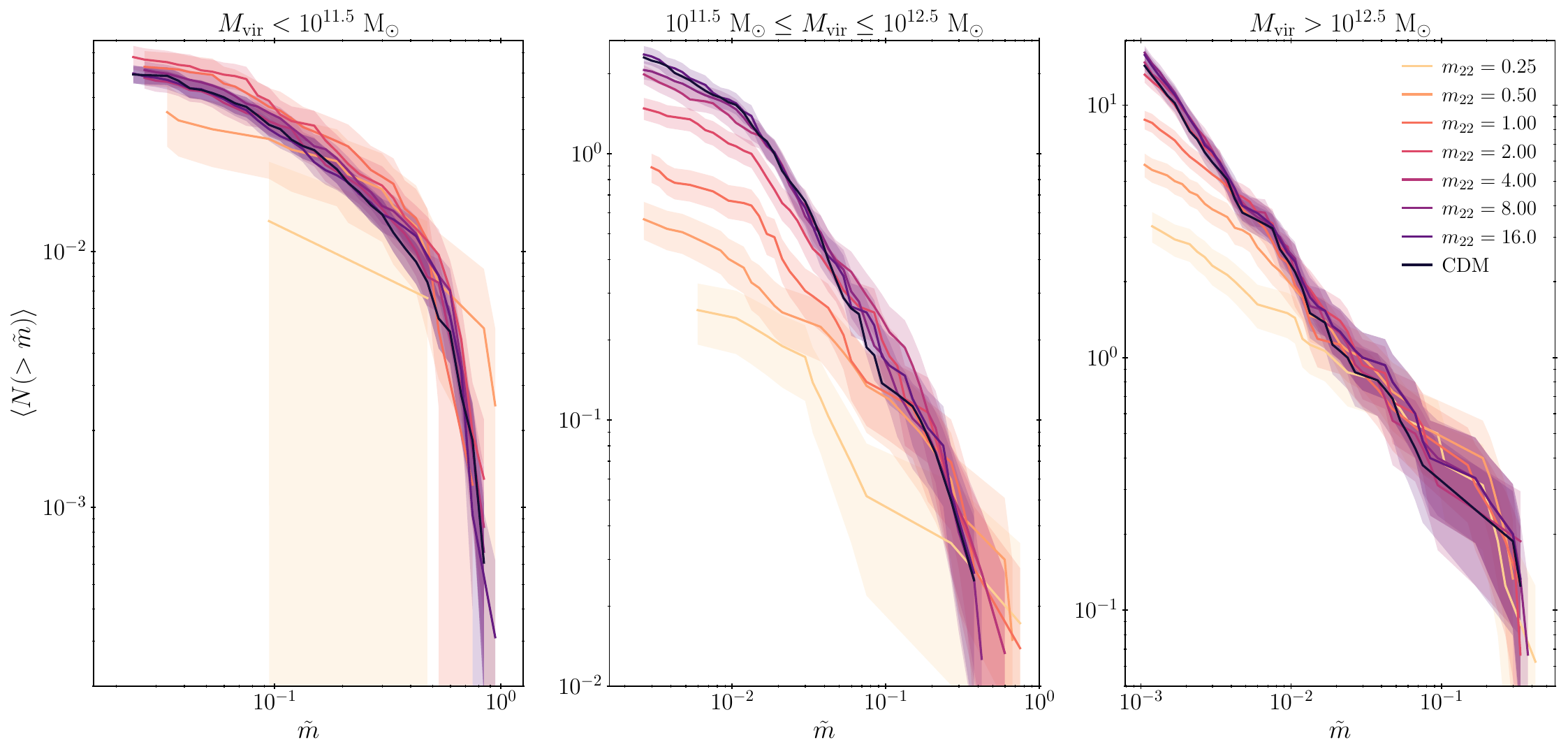}
\caption{The cumulative SHMFs averaged over all systems with $M_{\text{vir}} < 10^{11.5}$ M$_{\sun}$ (left plot),  $ 10^{11.5}$ M$_{\sun} <  M_{\text{vir}} < 10^{12.5}$ M$_{\sun}$ (middle plot), and $M_{\text{vir}} > 10^{12.5}$ M$_{\sun}$ (right plot). The $1 \sigma$ statistical errors are included as shaded regions.}
\label{fig:shmf}
\end{figure*}

Fig.~\ref{fig:shmf} shows the SHMFs for CDM and FDM with different FDM particle masses for the three mass bins. The left plot represents systems within bin $1$, the middle plot corresponds to systems within the MW-like bin, and the right plot is for systems within bin $3$. For bin 3 (the highest mass bin), we observe a suppression at low $\Tilde{m}$, which indicates that these large systems suffer an increasing lack of the smallest satellites as the FDM mass decreases. Likewise, the MW-like (middle mass) bin exhibits a similar trend, with the impact of FDM extending to the largest satellites for the lowest FDM masses. Notably, bin 1 (the lowest mass bin) appears to show no significant deviation. This is because the suppression due to FDM is so strong that it impacts the mass scales of the host haloes, and hence those affected systems are completely absent. Similar to the pattern observed in the HMF, we find that the suppression of substructure is stronger for lower FDM masses.

\begin{figure*}
\includegraphics[width=\textwidth, trim={0.16cm 0.11cm 0.11cm 0.11cm}, clip] {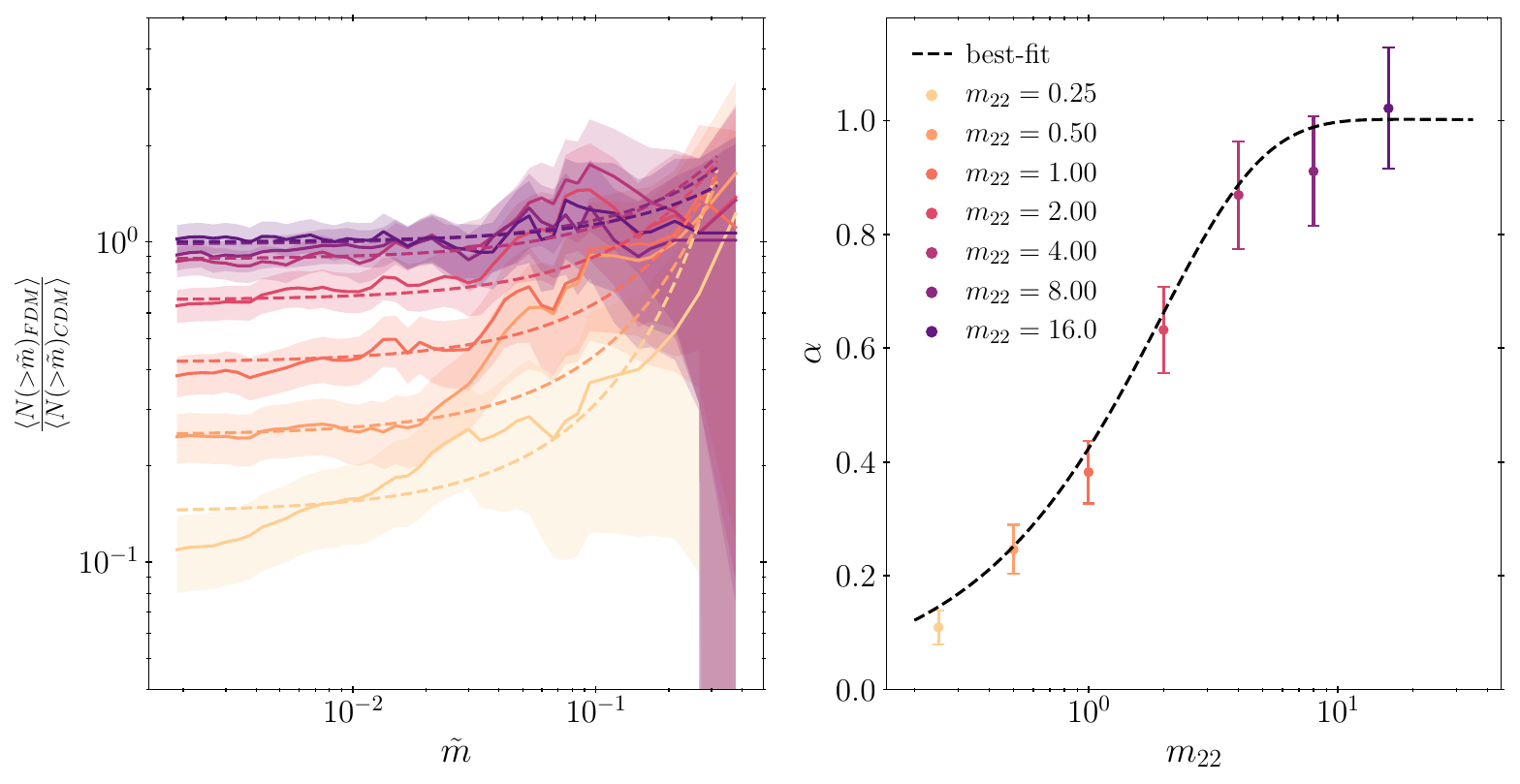}
\caption{The left plot shows the FDM-to-CDM cumulative subhalo abundance ratio averaged over systems within the MW-like mass bin for different FDM masses. The dashed curves show this ratio computed using the fitting formula provided in this work which is given by Eq.~\ref{eq:SHMF_fit}. The $1 \sigma$ statistical errors are included as shaded regions. The right plot (which is a subset of the left plot) shows the averaged FDM-to-CDM cumulative subhalo abundance ratio computed at $\mratio \approx 10^{-3}$ as a function of the FDM mass. The dashed black curve is again the best-fitting result using Eq.~\ref{eq:SHMF_fit}. }.
\label{fig:shmf_MW_bin}
\end{figure*}

For the MW-like bin, we provide a fitting formula for the FDM-to-CDM subhalo abundance ratio which takes the following analytical form:
\begin{equation}
\label{eq:SHMF_fit}
\frac{\langle N(>\mratio)_{FDM} \rangle}{\langle N(>\mratio)_{CDM} \rangle}= \tanh{(\kappa  m_{22}^{\beta})} e^{\mu {m_{22}^{-\gamma}} \mratio},
\end{equation}
where $\kappa, \beta, \mu$ and $\gamma$ are free parameters which we fit using Markov Chain Monte Carlo (MCMC) technique utilizing the MCMC python package \textsc{emcee} \citep{foreman2013emcee}. Using flat priors, we obtain the following mean posterior values for the parameters $\kappa = 0.448$, $\beta = 0.817$, $\mu = 4.27$ and $\gamma = 0.434$ from the MCMC run. The left plot of Fig.~\ref{fig:shmf_MW_bin} shows the FDM-to-CDM cumulative subhalo abundance ratio as obtained from the simulations (solid curves), as well as their respective fits (dashed curves) which were computed using Eq.~\ref{eq:SHMF_fit}. 

To quantify the suppression of substructure abundance for different FDM masses with respect to CDM, we define the parameter $\alpha$ as:
\begin{equation}
\label{eq:alpha}
\alpha \equiv \left. \frac{\langle N(>\mratio)_{FDM} \rangle}{\langle N(>\mratio)_{CDM} \rangle} \right|_{\mratio \approx 10^{-3}},
\end{equation}
which represents the FDM-to-CDM ratio of the average number of subhaloes residing in MW-like systems and that have masses larger than $\approx 1/1000$ of their host virial mass. Thus, this new variable $\alpha$ captures the departure of FDM models from CDM in terms of substructure abundance suppression in MW-like systems (where by definition, $\alpha = 1$ for a particular FDM model indicates that the FDM subhalo abundance identically matches that of CDM). To quantify the substructure suppression in FDM, the right plot of Fig.~\ref{fig:shmf_MW_bin} shows $\alpha$ as a function of the FDM particle mass. The dashed black curve is the best-fit which was computed using Eq. \ref{eq:SHMF_fit} at $\mratio \approx 10^{-3}$. For MW-like systems, this can be used to deduce that the satellite counts of $m_{22} \approx 0.1$ and $m_{22} \approx 1$ are $10\%$ and $50\%$ relative to CDM, respectively.

\subsection{Properties of density cores}
\label{subsec:density_cores_properties}

\begin{figure*}
\includegraphics[width=500pt, trim={0.16cm 0.11cm 0.11cm 0.11cm}, clip] {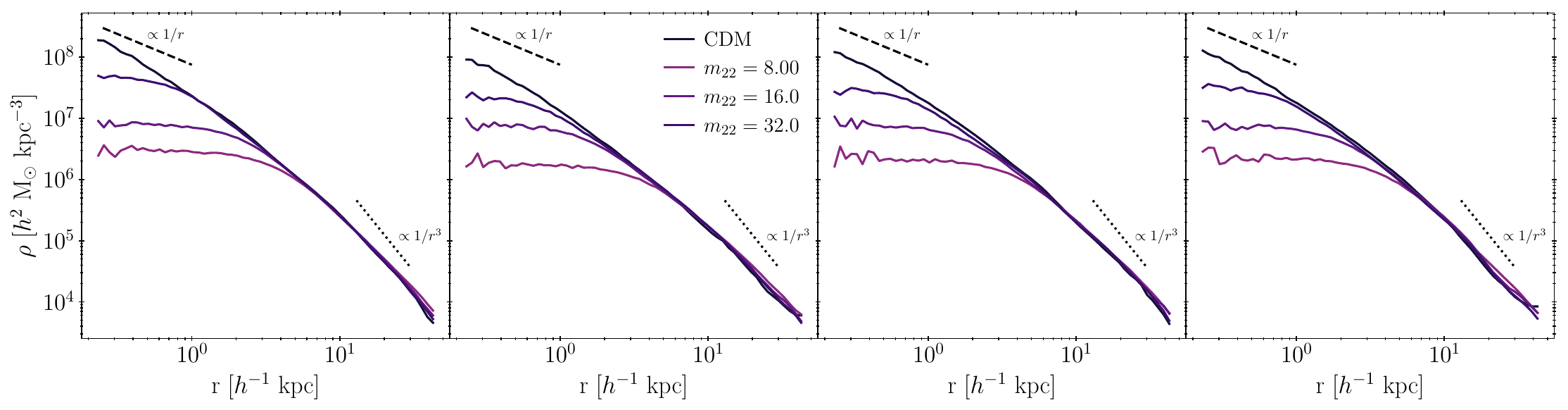}
\caption{Radial density profiles for CDM and FDM with masses $m_{22} \in \{8.00, 16.0, 32.0\}$ for systems A, B, C, and D from left to right. The dashed and dotted black lines show the NFW profile scalings of $\propto r^{-1}$ and $\propto r^{-3}$, respectively.}
\label{fig:density_prof}
\end{figure*}

The radial density profiles of CDM and FDM subhaloes with particle masses $m_{22} \in \{8.00, 16.0, 32.0\}$ are measured from the zoom-in high-resolution simulations for systems A, B, C, and D, as presented in Fig.~\ref{fig:density_prof}. The expected scalings for the NFW profile, characterized by a logarithmic inner slope of $-1$ representing the central cusp and a logarithmic outer slope of $-3$ representing the NFW envelope, are shown by the dashed and dotted black lines, respectively. As expected, the FDM subhaloes exhibit flat inner density profiles, indicating significant core formation, while the CDM subhaloes exhibit the expected central cusps. 

From Fig.~\ref{fig:density_prof}, it can be observed that significant cores are able to form in dwarf galaxies with mass of $\mathcal{O}(10^{9}$ M$_{\sun})$ in an FDM universe. This is important to emphasize since star formation has been shown to be insufficient to perturb the dark matter and induce the cusp-core transformation inside haloes with a mass below $5\times 10^9$ M$_{\sun}$ \citep{cintio2014mass, chan2015impact, 2016MNRAS.456.3542T}, given the very limited contribution of stars and gas to the whole potential \citep[e.g.][]{Maccio2020}.

To extract the FDM core radii from the radial profiles in Fig.~\ref{fig:density_prof}, we employ the cored-NFW density profile formula given by Eq.~\ref{eq:cored-NFW}. By imposing continuity at $r = r_{t}$, the number of free parameters are reduced by one to $\rho_c, r_s, r_c, r_t$. For each system, those four parameters are fitted by performing an MCMC exploration of the parameter space using the python package \textsc{emcee} \citep{foreman2013emcee}.

For completeness, all the density profile parameter values $r_{c}, \rho_{c}, r_{s}, r_{t}$, and $m_{22}$ for all four systems are given in Table~\ref{tab:dens_core_properties}. The density profile parameter values reported for each FDM mass are the median posterior distribution values from the MCMC run, while the mean parameter values are derived by averaging over all four systems. For the FDM masses $m_{22} \in \{8.00, 16.0, 32.0\}$, we find the mean core radii to be: $\langle r_{c} \rangle = 1.75 \pm 0.07$ $h^{-1}$ kpc, $\langle r_{c} \rangle = 0.88 \pm 0.04$ $h^{-1}$ kpc, and $\langle r_{c} \rangle = 0.45 \pm 0.02$ $h^{-1}$ kpc, respectively. 

\begin{table*}
\centering
\begin{tabular}{ c  c  c  c  c  c } 
\hline
$m_{22}$ & System & $r_{c}$ & $\log {(\rho_{c}/ h^{2} \mathrm{M}_{\sun} \mathrm{kpc}^{-3})}$  & $r_{s}$ & $r_{t}$ \\
& &  ($h^{-1}$ kpc) & & ($h^{-1}$ kpc) & ($h^{-1}$ kpc) \\\hline
 & A & 1.66 & 7.16 & 2.88 & 1.98 \\ 
8.00 & B & 1.75 & 6.92 & 5.11 & 2.61 \\ 
 & C & 1.77 & 7.02 & 3.36 & 2.07 \\
& D & 1.77 & 7.03 & 2.71 & 2.15 \\ \hline
 & A & 1.02 & 7.59 & 3.42 & 1.81 \\
16.0 & B & 0.807 & 7.54 & 3.34 & 1.18 \\ 
 & C & 0.794 & 7.59 & 3.10 & 0.824 \\ 
& D & 0.895 & 7.57 & 2.56 & 0.976 \\ \hline
  & A & 0.466 & 8.29 & 2.38 & 0.843 \\
32.0 & B & 0.425 & 8.02 & 2.84 & 0.648 \\
 & C & 0.456 & 8.11 & 2.29 & 0.639 \\
& D & 0.460 & 8.14 & 3.05 & 0.843 \\ \hline
 8.00 &   & 1.74 & 7.03 & 3.52 & 2.20 \\ 
 16.0 & mean & 0.880 & 7.57 & 3.10 & 1.20 \\ 
 32.0 &  & 0.451 & 8.14 & 2.64 & 0.744 \\
\hline
\end{tabular}
\caption{\label{tab:widgets}Density profile parameters for systems A, B, C, and D with particle masses $m_{22} \in \{8.00, 16.0, 32.0\}$, along with their mean values. The quoted parameter values are the median of the posterior distribution found from the MCMC run.}
\label{tab:dens_core_properties}
\end{table*}

\begin{figure*}
    \centering
    \includegraphics[width=\textwidth]{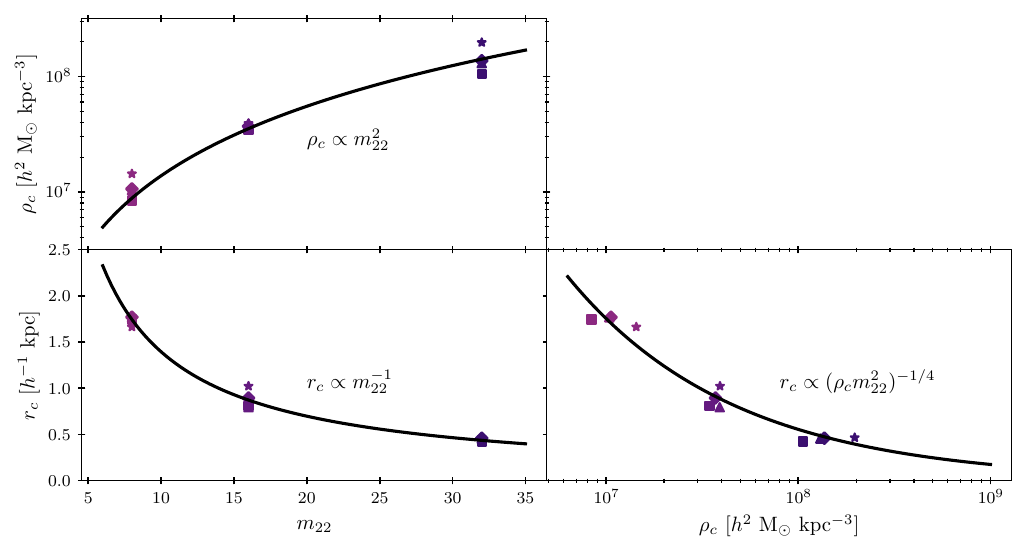}
    \caption{The FDM core-scaling relations. Upper and lower left plots show the core density and radius verses the FDM particle mass, respectively, while the lower right plot shows the core radius as a function of the core density. Systems A, B, C, and D are denoted by star, square, triangle, and diamond, respectively. The black curves show the respective best-fitting scaling relations.}
    \label{fig:scaling_rel}
\end{figure*}

We investigate the dependence of the core properties given in Table~\ref{tab:dens_core_properties}~--~namely the core radius and density~--~on the FDM particle mass. Fig.~\ref{fig:scaling_rel} presents the relations between $r_{c}, \rho_{c},$ and $m_{22}$ for all four systems: $r_{c} \propto (m_{22}^2 \rho_{c})^{-1/4}$, $r_{c} \propto m_{22}^{-1}$, and $\rho_{c} \propto m_{22}^2$. By extrapolating the core radius-FDM mass relation $(r_{c} \propto m_{22}^{-1})$ to lower masses, we are able to estimate the core radii for the FDM masses $m_{22} = \{0.25, 0.50, 1.00, 2.00, 4.00\}$ for systems which were not simulated in this work (as those specific systems were not present).

\subsection{Catch-22 or not Catch-22}

As a final step, we want to determine if there exists a {\it Catch-22} problem for FDM~--~namely, if there is a region in the parameter space where one can simulteously have enough dark matter haloes to host the observed number of satellites around the MW and M31 while still form substantial cores in the dark matter distribution.

Given the inefficiency of star formation at low halo masses \citep[e.g.][]{Bullock2000, Behroozi2013, Buck2019}, one needs to have a substantial population of dark matter haloes to act as possible cradles for the formation of dwarf satellite galaxies. In line with previous studies based on other dark matter models \citep[][]{MaccioFontanot10, Lovell2012, Lovell13}, we exclude the region where $\alpha < 0.5$. This choice is quite arbitrary and mainly serves an illustrative purpose, demonstrating that substantial cores are formed even for more conservative FDM models. In fact, this is quite a conservative choice given that even models with a lower value of $\alpha$ might still be able to host the 30-40 satellites observed around our own galaxy. We plan to revisit this threshold once we have full hydrodynamical simulations in the FDM model.
As will be discussed later, even though the constraint on $\alpha$ is only a rough one, Fig.~\ref{fig:shmf_MW_bin} illustrates that the corresponding lower-limit on the FDM mass is quite insensitive to this $\alpha$ threshold.

On the density profiles side, we exclude $r_{c} > 2$ $h^{-1}$ kpc. Such large cores will be incompatible with current observations of the matter distribution in the most studied satellites on the MW, namely Carina, Fornax, Sextans and Sculptor \citep[e.g.][]{Read2019, Marquez2023}. From the $r_{c}-m_{22}$ relation shown in Fig.~\ref{fig:scaling_rel}, excluding $r_{c} > 2$ $h^{-1}$ kpc corresponds to excluding $m_{22} \lesssim 8$.

\begin{figure*}
    \includegraphics[width=\textwidth]{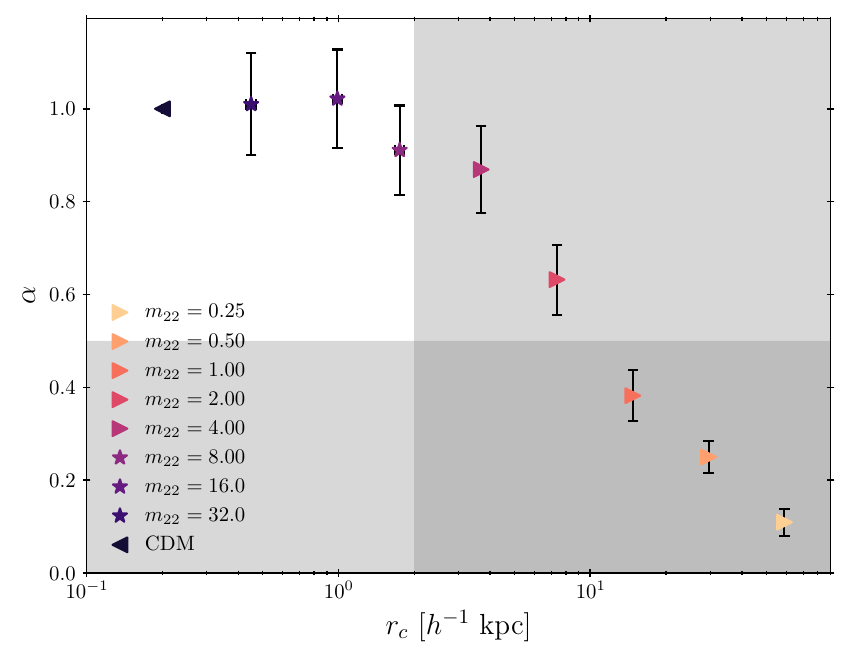}
    \caption{The FDM-to-CDM satellite abundance reduction ratio verses the core radius for all the FDM masses. The core radii for the masses $m_{22} \in \{0.25, 0.50, 1.00, 2.00, 4.00\}$ were found via extrapolation of the fitted core radius-mass scaling relation provided in Fig.~\ref{fig:scaling_rel}. The grey regions corresponding to $\alpha < 0.5$ and $r_{c} > 2$ $h^{-1}$ kpc are the excluded regions. The occupancy of the top-left unshaded region by FDM masses illustrates the absence of a \textit{Catch-22} problem in an FDM universe.}
    \label{fig:alpha_rc}
\end{figure*}

Fig.~\ref{fig:alpha_rc} shows the $\alpha-r_{c}$ parameter space for the different FDM masses. The shaded regions correspond to the excluded regions in this parameter space (namely, $\alpha < 0.5$ and $r_{c} > 2$ $h^{-1}$ kpc). The core radii for the masses $m_{22} = \{0.25, 0.50, 1.00, 2.00, 4.00\}$ were estimated using the $r_{c}-m_{22}$ relation. The value of $\alpha$ for $m_{22} = 32$ was estimated using the fitting formula given by Eq.~\ref{eq:SHMF_fit}.  The occupancy of the top-left unshaded region by FDM masses illustrates that it is possible to find a FDM particle mass interval --~namely $m_{22} \gtrsim 8$~-- where FDM produces both the observed substructure counts and core sizes at the same time. We therefore conclude that FDM does not undergo a \textit{Catch-22} problem. 

The simulations used in this work do not incorporate
the effect of baryons on the dark matter. While baryonic physics is known to have non-trivial impact on the dark matter distribution on sub-galactic scale, it is important to stress that our conclusion that there is no \textit{Catch-22} problem in FDM is expected to still hold even with the addition of baryons. This is illustrated by Fig.~\ref{fig:alpha_rc}, where no matter how one shifts the $\alpha$ and/or $r_{c}$ thresholds, there will \textit{always} exist an unshaded region in the $\alpha-r_{c}$ parameter space that is occupied by FDM masses. In addition, the derived FDM mass constraint after the inclusion of baryons is not expected to be significantly different as well. This is the case because on the substructure abundance side, the FDM mass is quite insensitive to changes in $\alpha$ (refer to Fig.~\ref{fig:shmf_MW_bin}). This means that large changes in $\alpha$ translate to only marginal changes in the FDM mass constraint. Furthermore, as previously discussed on the density profile front, the fact that we examined dwarf galaxies ($\approx 10^{9}-10^{10}$ M$_{\sun}$) which are known to have only minimal baryonic content ensures that the core radii of such systems will not significantly change with the introduction of baryons.

\section{Conclusions}
\label{sec:conclusions}
We presented the FDM SubHalo Mass Function (SHMF) as measured from cosmological volume simulations which fully model the effects of the quantum potential throughout cosmic evolution for seven different FDM particle masses in the range $m_{22} \in [0.25 - 16]$. In addition, we computed the density profiles from zoom-in high-resolution simulations of dwarf galaxies for the FDM masses in the overlapping range $m_{22} \in [8 - 32]$, and used those to infer the size of the solitonic core within each dwarf galaxy.  We summarize our main findings below:
\begin{itemize}
  \item For the first time, we present a fitting functional form for the cumulative FDM-to-CDM SHMF ratio using FDM simulations that take full account of the dynamical effects of the quantum potential throughout cosmic evolution;
  \item In an FDM universe, we find that it is possible to form significant cores in dwarf galaxies ($\approx 10^9$ M$_{\sun}$). This was reported to be very difficult to achieve with baryonic feedback alone;
  \item Unlike WDM, we conclude that FDM does not undergo a \textit{Catch-22} problem. In particular, one can have particle masses ($m_{22} \gtrsim 8$) that simultaneously produce both substantial cores and the observed satellite abundances.  
\end{itemize}

\section*{Acknowledgements}
This material is based upon work supported by Tamkeen under the New York University Abu Dhabi Research Institute grant CASS. The authors gratefully acknowledge the High Performance Computing resources at New York University Abu Dhabi. MB is supported by the project ``Combining Cosmic Microwave Background and Large Scale Structure data: an Integrated Approach for Addressing Fundamental Questions in Cosmology'', funded by the MIUR Progetti di Ricerca di Rilevante Interesse Nazionale (PRIN) Bando 2017 - grant 2017YJYZAH.

\section*{Data Availability}
The data underlying this article will be shared on reasonable request to the corresponding author.

%%%%%%%%%%%%%%%%%%%% REFERENCES %%%%%%%%%%%%%%%%%%

\bibliographystyle{mnras}
\bibliography{BIB,baldi_bibliography} 

\begin{thebibliography}{}
\makeatletter
\relax
\def\mn@urlcharsother{\let\do\@makeother \do\$\do\&\do\#\do\^\do\_\do\%\do\~}
\def\mn@doi{\begingroup\mn@urlcharsother \@ifnextchar [ {\mn@doi@}
  {\mn@doi@[]}}
\def\mn@doi@[#1]#2{\def\@tempa{#1}\ifx\@tempa\@empty \href
  {http://dx.doi.org/#2} {doi:#2}\else \href {http://dx.doi.org/#2} {#1}\fi
  \endgroup}
\def\mn@eprint#1#2{\mn@eprint@#1:#2::\@nil}
\def\mn@eprint@arXiv#1{\href {http://arxiv.org/abs/#1} {{\tt arXiv:#1}}}
\def\mn@eprint@dblp#1{\href {http://dblp.uni-trier.de/rec/bibtex/#1.xml}
  {dblp:#1}}
\def\mn@eprint@#1:#2:#3:#4\@nil{\def\@tempa {#1}\def\@tempb {#2}\def\@tempc
  {#3}\ifx \@tempc \@empty \let \@tempc \@tempb \let \@tempb \@tempa \fi \ifx
  \@tempb \@empty \def\@tempb {arXiv}\fi \@ifundefined
  {mn@eprint@\@tempb}{\@tempb:\@tempc}{\expandafter \expandafter \csname
  mn@eprint@\@tempb\endcsname \expandafter{\@tempc}}}

\bibitem[\protect\citeauthoryear{Abazajian}{Abazajian}{2006}]{abazajian2006production}
Abazajian K.,  2006, Physical Review D, 73, 063506

\bibitem[\protect\citeauthoryear{Aghanim et~al.,}{Aghanim
  et~al.}{2020}]{aghanim2020planck}
Aghanim N.,  et~al., 2020, Astronomy \& Astrophysics, 641, A6

\bibitem[\protect\citeauthoryear{{Alam} et~al.,}{{Alam}
  et~al.}{2017}]{2017MNRAS.470.2617A}
{Alam} S.,  et~al., 2017, \mn@doi [\mnras] {10.1093/mnras/stx721}, \href
  {https://ui.adsabs.harvard.edu/abs/2017MNRAS.470.2617A} {470, 2617}

\bibitem[\protect\citeauthoryear{{Amendola} \& {Barbieri}}{{Amendola} \&
  {Barbieri}}{2006}]{Amendola2006}
{Amendola} L.,  {Barbieri} R.,  2006, \mn@doi [Physics Letters B]
  {10.1016/j.physletb.2006.08.069}, \href
  {https://ui.adsabs.harvard.edu/abs/2006PhLB..642..192A} {642, 192}

\bibitem[\protect\citeauthoryear{Amorisco \& Loeb}{Amorisco \&
  Loeb}{2018}]{amorisco2018first}
Amorisco N.~C.,  Loeb A.,  2018, arXiv preprint arXiv:1808.00464

\bibitem[\protect\citeauthoryear{Bahcall, Lubin  \& Dorman}{Bahcall
  et~al.}{1995}]{bahcall1995dark}
Bahcall N.~A.,  Lubin L.~M.,   Dorman V.,  1995, The Astrophysical Journal,
  447, L81

\bibitem[\protect\citeauthoryear{{Behroozi}, {Wechsler}  \&
  {Conroy}}{{Behroozi} et~al.}{2013}]{Behroozi2013}
{Behroozi} P.~S.,  {Wechsler} R.~H.,   {Conroy} C.,  2013, \mn@doi [\apj]
  {10.1088/0004-637X/770/1/57}, \href
  {https://ui.adsabs.harvard.edu/abs/2013ApJ...770...57B} {770, 57}

\bibitem[\protect\citeauthoryear{Bosma}{Bosma}{1981}]{bosma198121}
Bosma A.,  1981, The Astronomical Journal, 86, 1825

\bibitem[\protect\citeauthoryear{Brooks \& Zolotov}{Brooks \&
  Zolotov}{2014}]{brooks2014baryons}
Brooks A.~M.,  Zolotov A.,  2014, The Astrophysical Journal, 786, 87

\bibitem[\protect\citeauthoryear{{Buck}, {Macci{\`o}}, {Dutton}, {Obreja}  \&
  {Frings}}{{Buck} et~al.}{2019}]{Buck2019}
{Buck} T.,  {Macci{\`o}} A.~V.,  {Dutton} A.~A.,  {Obreja} A.,   {Frings} J.,
  2019, \mn@doi [\mnras] {10.1093/mnras/sty2913}, \href
  {https://ui.adsabs.harvard.edu/abs/2019MNRAS.483.1314B} {483, 1314}

\bibitem[\protect\citeauthoryear{{Bullock}, {Kravtsov}  \&
  {Weinberg}}{{Bullock} et~al.}{2000}]{Bullock2000}
{Bullock} J.~S.,  {Kravtsov} A.~V.,   {Weinberg} D.~H.,  2000, \mn@doi [\apj]
  {10.1086/309279}, \href
  {https://ui.adsabs.harvard.edu/abs/2000ApJ...539..517B} {539, 517}

\bibitem[\protect\citeauthoryear{Chan, Kere{\v{s}}, O{\~n}orbe, Hopkins,
  Muratov, Faucher-Giguere  \& Quataert}{Chan et~al.}{2015}]{chan2015impact}
Chan T.,  Kere{\v{s}} D.,  O{\~n}orbe J.,  Hopkins P.,  Muratov A.,
  Faucher-Giguere C.-A.,   Quataert E.,  2015, Monthly Notices of the Royal
  Astronomical Society, 454, 2981

\bibitem[\protect\citeauthoryear{Chan, Ferreira, May, Hayashi  \& Chiba}{Chan
  et~al.}{2022}]{chan2022diversity}
Chan H. Y.~J.,  Ferreira E.~G.,  May S.,  Hayashi K.,   Chiba M.,  2022,
  Monthly Notices of the Royal Astronomical Society, 511, 943

\bibitem[\protect\citeauthoryear{{Chavanis}}{{Chavanis}}{2011}]{Chavanis11a}
{Chavanis} P.-H.,  2011, \mn@doi [\prd] {10.1103/PhysRevD.84.043531}, \href
  {https://ui.adsabs.harvard.edu/abs/2011PhRvD..84d3531C} {84, 043531}

\bibitem[\protect\citeauthoryear{Child, Habib, Heitmann, Frontiere, Finkel,
  Pope  \& Morozov}{Child et~al.}{2018}]{child2018halo}
Child H.~L.,  Habib S.,  Heitmann K.,  Frontiere N.,  Finkel H.,  Pope A.,
  Morozov V.,  2018, The Astrophysical Journal, 859, 55

\bibitem[\protect\citeauthoryear{Copi, Schramm  \& Turner}{Copi
  et~al.}{1995}]{copi1995big}
Copi C.~J.,  Schramm D.~N.,   Turner M.~S.,  1995, Science, 267, 192

\bibitem[\protect\citeauthoryear{{Davis}, {Efstathiou}, {Frenk}  \&
  White}{{Davis} et~al.}{1985}]{Davis_etal_1985}
{Davis} M.,  {Efstathiou} G.,  {Frenk} C.~S.,   White S.~D.,  1985, \mn@doi
  [Astrophys.J.] {10.1086/163168}, 292, 371

\bibitem[\protect\citeauthoryear{Di~Cintio, Brook, Dutton, Macci{\`o}, Stinson
  \& Knebe}{Di~Cintio et~al.}{2014}]{cintio2014mass}
Di~Cintio A.,  Brook C.~B.,  Dutton A.~A.,  Macci{\`o} A.~V.,  Stinson G.~S.,
  Knebe A.,  2014, Monthly Notices of the Royal Astronomical Society, 441, 2986

\bibitem[\protect\citeauthoryear{{Dutton} \& {Macci{\`o}}}{{Dutton} \&
  {Macci{\`o}}}{2014}]{dutton2014cold}
{Dutton} A.~A.,  {Macci{\`o}} A.~V.,  2014, \mn@doi [\mnras]
  {10.1093/mnras/stu742}, \href
  {https://ui.adsabs.harvard.edu/abs/2014MNRAS.441.3359D} {441, 3359}

\bibitem[\protect\citeauthoryear{{Exp{\'o}sito-M{\'a}rquez}, {Brook},
  {Huertas-Company}, {Di Cintio}, {Macci{\`o}}, {Grand}, {Battaglia}  \&
  {Arjona-G{\'a}lvez}}{{Exp{\'o}sito-M{\'a}rquez} et~al.}{2023}]{Marquez2023}
{Exp{\'o}sito-M{\'a}rquez} J.,  {Brook} C.~B.,  {Huertas-Company} M.,  {Di
  Cintio} A.,  {Macci{\`o}} A.~V.,  {Grand} R.~J.~J.,  {Battaglia} G.,
  {Arjona-G{\'a}lvez} E.,  2023, \mn@doi [\mnras] {10.1093/mnras/stac3799},
  \href {https://ui.adsabs.harvard.edu/abs/2023MNRAS.519.4384E} {519, 4384}

\bibitem[\protect\citeauthoryear{Ferreira}{Ferreira}{2021}]{ferreira2021ultra}
Ferreira E.~G.,  2021, The Astronomy and Astrophysics Review, 29, 1

\bibitem[\protect\citeauthoryear{{Fitts} et~al.,}{{Fitts}
  et~al.}{2017}]{2017MNRAS.471.3547F}
{Fitts} A.,  et~al., 2017, \mn@doi [\mnras] {10.1093/mnras/stx1757}, \href
  {https://ui.adsabs.harvard.edu/abs/2017MNRAS.471.3547F} {471, 3547}

\bibitem[\protect\citeauthoryear{Foreman-Mackey, Hogg, Lang  \&
  Goodman}{Foreman-Mackey et~al.}{2013}]{foreman2013emcee}
Foreman-Mackey D.,  Hogg D.~W.,  Lang D.,   Goodman J.,  2013, Publications of
  the Astronomical Society of the Pacific, 125, 306

\bibitem[\protect\citeauthoryear{{Frings}, {Macci{\`o}}, {Buck}, {Penzo},
  {Dutton}, {Blank}  \& {Obreja}}{{Frings} et~al.}{2017}]{Frings2017}
{Frings} J.,  {Macci{\`o}} A.,  {Buck} T.,  {Penzo} C.,  {Dutton} A.,  {Blank}
  M.,   {Obreja} A.,  2017, \mn@doi [\mnras] {10.1093/mnras/stx2171}, \href
  {https://ui.adsabs.harvard.edu/abs/2017MNRAS.472.3378F} {472, 3378}

\bibitem[\protect\citeauthoryear{Gilman, Du, Benson, Birrer, Nierenberg  \&
  Treu}{Gilman et~al.}{2020b}]{gilman2020constraints}
Gilman D.,  Du X.,  Benson A.,  Birrer S.,  Nierenberg A.,   Treu T.,  2020b,
  Monthly Notices of the Royal Astronomical Society: Letters, 492, L12

\bibitem[\protect\citeauthoryear{{Gilman}, {Du}, {Benson}, {Birrer},
  {Nierenberg}  \& {Treu}}{{Gilman} et~al.}{2020a}]{2020MNRAS.492L..12G}
{Gilman} D.,  {Du} X.,  {Benson} A.,  {Birrer} S.,  {Nierenberg} A.,   {Treu}
  T.,  2020a, \mn@doi [\mnras] {10.1093/mnrasl/slz173}, \href
  {https://ui.adsabs.harvard.edu/abs/2020MNRAS.492L..12G} {492, L12}

\bibitem[\protect\citeauthoryear{Hlozek, Grin, Marsh  \& Ferreira}{Hlozek
  et~al.}{2015}]{axionCAMB}
Hlozek R.,  Grin D.,  Marsh D. J.~E.,   Ferreira P.~G.,  2015, \mn@doi [Phys.
  Rev.] {10.1103/PhysRevD.91.103512}, D91, 103512

\bibitem[\protect\citeauthoryear{Hu, Barkana  \& Gruzinov}{Hu
  et~al.}{2000}]{Hu00}
Hu W.,  Barkana R.,   Gruzinov A.,  2000, \mn@doi [Phys. Rev. Lett.]
  {10.1103/PhysRevLett.85.1158}, 85, 1158

\bibitem[\protect\citeauthoryear{Hui}{Hui}{2021}]{hui2021wave}
Hui L.,  2021, Annual Review of Astronomy and Astrophysics, 59, 247

\bibitem[\protect\citeauthoryear{Hui, Ostriker, Tremaine  \& Witten}{Hui
  et~al.}{2017a}]{PhysRevD.95.043541}
Hui L.,  Ostriker J.~P.,  Tremaine S.,   Witten E.,  2017a, \mn@doi [Phys. Rev.
  D] {10.1103/PhysRevD.95.043541}, 95, 043541

\bibitem[\protect\citeauthoryear{Hui, Ostriker, Tremaine  \& Witten}{Hui
  et~al.}{2017b}]{Hui16}
Hui L.,  Ostriker J.~P.,  Tremaine S.,   Witten E.,  2017b, \mn@doi [Phys.
  Rev.] {10.1103/PhysRevD.95.043541}, D95, 043541

\bibitem[\protect\citeauthoryear{Ji \& Sin}{Ji \& Sin}{1994}]{Ji94}
Ji S.~U.,  Sin S.~J.,  1994, \mn@doi [Phys. Rev.] {10.1103/PhysRevD.50.3655},
  D50, 3655

\bibitem[\protect\citeauthoryear{{Komatsu} et~al.,}{{Komatsu}
  et~al.}{2011}]{2011ApJS..192...18K}
{Komatsu} E.,  et~al., 2011, \mn@doi [\apjs] {10.1088/0067-0049/192/2/18},
  \href {https://ui.adsabs.harvard.edu/abs/2011ApJS..192...18K} {192, 18}

\bibitem[\protect\citeauthoryear{Kulkarni \& Ostriker}{Kulkarni \&
  Ostriker}{2022}]{kulkarni2022halo}
Kulkarni M.,  Ostriker J.~P.,  2022, Monthly Notices of the Royal Astronomical
  Society, 510, 1425

\bibitem[\protect\citeauthoryear{{Lovell} et~al.,}{{Lovell}
  et~al.}{2012}]{Lovell2012}
{Lovell} M.~R.,  et~al., 2012, \mn@doi [\mnras]
  {10.1111/j.1365-2966.2011.20200.x}, \href
  {https://ui.adsabs.harvard.edu/abs/2012MNRAS.420.2318L} {420, 2318}

\bibitem[\protect\citeauthoryear{Lovell, Frenk, Eke, Jenkins, Gao  \&
  Theuns}{Lovell et~al.}{2014a}]{Lovell13}
Lovell M.~R.,  Frenk C.~S.,  Eke V.~R.,  Jenkins A.,  Gao L.,   Theuns T.,
  2014a, \mn@doi [Mon. Not. Roy. Astron. Soc.] {10.1093/mnras/stt2431}, 439,
  300

\bibitem[\protect\citeauthoryear{Lovell, Frenk, Eke, Jenkins, Gao  \&
  Theuns}{Lovell et~al.}{2014b}]{lovell2014properties}
Lovell M.~R.,  Frenk C.~S.,  Eke V.~R.,  Jenkins A.,  Gao L.,   Theuns T.,
  2014b, Monthly Notices of the Royal Astronomical Society, 439, 300

\bibitem[\protect\citeauthoryear{Ludlow, Bose, Angulo, Wang, Hellwing, Navarro,
  Cole  \& Frenk}{Ludlow et~al.}{2016}]{ludlow2016mass}
Ludlow A.~D.,  Bose S.,  Angulo R.~E.,  Wang L.,  Hellwing W.~A.,  Navarro
  J.~F.,  Cole S.,   Frenk C.~S.,  2016, Monthly Notices of the Royal
  Astronomical Society, 460, 1214

\bibitem[\protect\citeauthoryear{Macci{\`{o} } \& Fontanot}{Macci{\`{o} } \&
  Fontanot}{2010}]{MaccioFontanot10}
Macci{\`{o} } A.~V.,  Fontanot F.,  2010, \mn@doi [Monthly Notices of the Royal
  Astronomical Society: Letters] {10.1111/j.1745-3933.2010.00825.x}, 404, L16

\bibitem[\protect\citeauthoryear{{Macci{\`o}}, {Dutton}, {van den Bosch},
  {Moore}, {Potter}  \& {Stadel}}{{Macci{\`o}} et~al.}{2007}]{Maccio2007}
{Macci{\`o}} A.~V.,  {Dutton} A.~A.,  {van den Bosch} F.~C.,  {Moore} B.,
  {Potter} D.,   {Stadel} J.,  2007, \mn@doi [\mnras]
  {10.1111/j.1365-2966.2007.11720.x}, \href
  {https://ui.adsabs.harvard.edu/abs/2007MNRAS.378...55M} {378, 55}

\bibitem[\protect\citeauthoryear{{Macci{\`o}}, {Paduroiu}, {Anderhalden},
  {Schneider}  \& {Moore}}{{Macci{\`o}} et~al.}{2012}]{Maccio12}
{Macci{\`o}} A.~V.,  {Paduroiu} S.,  {Anderhalden} D.,  {Schneider} A.,
  {Moore} B.,  2012, \mn@doi [\mnras] {10.1111/j.1365-2966.2012.21284.x}, \href
  {https://ui.adsabs.harvard.edu/abs/2012MNRAS.424.1105M} {424, 1105}

\bibitem[\protect\citeauthoryear{{Macci{\`o}}, {Crespi}, {Blank}  \&
  {Kang}}{{Macci{\`o}} et~al.}{2020}]{Maccio2020}
{Macci{\`o}} A.~V.,  {Crespi} S.,  {Blank} M.,   {Kang} X.,  2020, \mn@doi
  [\mnras] {10.1093/mnrasl/slaa058}, \href
  {https://ui.adsabs.harvard.edu/abs/2020MNRAS.495L..46M} {495, L46}

\bibitem[\protect\citeauthoryear{Madelung}{Madelung}{1927}]{Madelung27}
Madelung E.,  1927, \mn@doi [Zeitschrift f{\"u}r Physik] {10.1007/BF01400372},
  40, 322

\bibitem[\protect\citeauthoryear{Marsh}{Marsh}{2016a}]{Marsh16nl}
Marsh D. J.~E.,  2016a, preprint, \href
  {http://adsabs.harvard.edu/abs/2016arXiv160505973M} {} (\mn@eprint {arXiv}
  {1605.05973})

\bibitem[\protect\citeauthoryear{Marsh}{Marsh}{2016b}]{MARSH20161}
Marsh D.~J.,  2016b, \mn@doi [Physics Reports]
  {https://doi.org/10.1016/j.physrep.2016.06.005}, 643, 1

\bibitem[\protect\citeauthoryear{{Navarro}, {Eke}  \& {Frenk}}{{Navarro}
  et~al.}{1996}]{Navarro1996}
{Navarro} J.~F.,  {Eke} V.~R.,   {Frenk} C.~S.,  1996, \mnras, \href
  {http://adsabs.harvard.edu/abs/1996MNRAS.283L..72N} {283, L72}

\bibitem[\protect\citeauthoryear{Navarro, Frenk  \& White}{Navarro
  et~al.}{1997}]{navarro1997universal}
Navarro J.~F.,  Frenk C.~S.,   White S.~D.,  1997, The Astrophysical Journal,
  490, 493

\bibitem[\protect\citeauthoryear{Nori \& Baldi}{Nori \& Baldi}{2018}]{Nori18}
Nori M.,  Baldi M.,  2018, \mn@doi [Mon. Not. Roy. Astron. Soc.]
  {10.1093/mnras/sty1224}, 478, 3935

\bibitem[\protect\citeauthoryear{Nori, Murgia, Ir\v{s}i\v{c}, Baldi  \&
  Viel}{Nori et~al.}{2019}]{Nori19}
Nori M.,  Murgia R.,  Ir\v{s}i\v{c} V.,  Baldi M.,   Viel M.,  2019, \mn@doi
  [Mon. Not. Roy. Astron. Soc.] {10.1093/mnras/sty2888}, 482, 3227

\bibitem[\protect\citeauthoryear{O{\~n}orbe, Boylan-Kolchin, Bullock, Hopkins,
  Kere{\v{s}}, Faucher-Gigu{\`e}re, Quataert  \& Murray}{O{\~n}orbe
  et~al.}{2015}]{onorbe2015forged}
O{\~n}orbe J.,  Boylan-Kolchin M.,  Bullock J.~S.,  Hopkins P.~F.,  Kere{\v{s}}
  D.,  Faucher-Gigu{\`e}re C.-A.,  Quataert E.,   Murray N.,  2015, Monthly
  Notices of the Royal Astronomical Society, 454, 2092

\bibitem[\protect\citeauthoryear{{Perlmutter} et~al.,}{{Perlmutter}
  et~al.}{1999}]{1999ApJ...517..565P}
{Perlmutter} S.,  et~al., 1999, \mn@doi [\apj] {10.1086/307221}, \href
  {https://ui.adsabs.harvard.edu/abs/1999ApJ...517..565P} {517, 565}

\bibitem[\protect\citeauthoryear{{Planck Collaboration} et~al.,}{{Planck
  Collaboration} et~al.}{2016}]{2016A&A...594A..13P}
{Planck Collaboration} et~al., 2016, \mn@doi [\aap]
  {10.1051/0004-6361/201525830}, \href
  {https://ui.adsabs.harvard.edu/abs/2016A&A...594A..13P} {594, A13}

\bibitem[\protect\citeauthoryear{{Read}, {Walker}  \& {Steger}}{{Read}
  et~al.}{2019}]{Read2019}
{Read} J.~I.,  {Walker} M.~G.,   {Steger} P.,  2019, \mn@doi [\mnras]
  {10.1093/mnras/sty3404}, \href
  {https://ui.adsabs.harvard.edu/abs/2019MNRAS.484.1401R} {484, 1401}

\bibitem[\protect\citeauthoryear{{Riess} et~al.,}{{Riess}
  et~al.}{1998}]{1998AJ....116.1009R}
{Riess} A.~G.,  et~al., 1998, \mn@doi [\aj] {10.1086/300499}, \href
  {https://ui.adsabs.harvard.edu/abs/1998AJ....116.1009R} {116, 1009}

\bibitem[\protect\citeauthoryear{{Rubin}, {Ford}  \& {Thonnard}}{{Rubin}
  et~al.}{1980}]{1980ApJ...238..471R}
{Rubin} V.~C.,  {Ford} W.~K. J.,   {Thonnard} N.,  1980, \mn@doi [\apj]
  {10.1086/158003}, \href
  {https://ui.adsabs.harvard.edu/abs/1980ApJ...238..471R} {238, 471}

\bibitem[\protect\citeauthoryear{Schive, Chiueh  \& Broadhurst}{Schive
  et~al.}{2014a}]{schive2014cosmic}
Schive H.-Y.,  Chiueh T.,   Broadhurst T.,  2014a, Nature Physics, 10, 496

\bibitem[\protect\citeauthoryear{Schive, Liao, Woo, Wong, Chiueh, Broadhurst
  \& Hwang}{Schive et~al.}{2014b}]{schive2014understanding}
Schive H.-Y.,  Liao M.-H.,  Woo T.-P.,  Wong S.-K.,  Chiueh T.,  Broadhurst T.,
    Hwang W.~P.,  2014b, Physical review letters, 113, 261302

\bibitem[\protect\citeauthoryear{Schive, Chiueh, Broadhurst  \& Huang}{Schive
  et~al.}{2016}]{schive2016contrasting}
Schive H.-Y.,  Chiueh T.,  Broadhurst T.,   Huang K.-W.,  2016, The
  Astrophysical Journal, 818, 89

\bibitem[\protect\citeauthoryear{{Schneider}, {Smith}, {Macci{\`o}}  \&
  {Moore}}{{Schneider} et~al.}{2012}]{Schneider2012}
{Schneider} A.,  {Smith} R.~E.,  {Macci{\`o}} A.~V.,   {Moore} B.,  2012,
  \mn@doi [\mnras] {10.1111/j.1365-2966.2012.21252.x}, \href
  {https://ui.adsabs.harvard.edu/abs/2012MNRAS.424..684S} {424, 684}

\bibitem[\protect\citeauthoryear{{Springel} et~al.,}{{Springel}
  et~al.}{2008a}]{Aquarius}
{Springel} V.,  et~al., 2008a, \mn@doi [\mnras]
  {10.1111/j.1365-2966.2008.14066.x}, \href
  {http://adsabs.harvard.edu/abs/2008MNRAS.391.1685S} {391, 1685}

\bibitem[\protect\citeauthoryear{Springel et~al.,}{Springel
  et~al.}{2008b}]{springel2008aquarius}
Springel V.,  et~al., 2008b, Monthly Notices of the Royal Astronomical Society,
  391, 1685

\bibitem[\protect\citeauthoryear{Tegmark et~al.,}{Tegmark
  et~al.}{2006}]{PhysRevD.74.123507}
Tegmark M.,  et~al., 2006, \mn@doi [Phys. Rev. D] {10.1103/PhysRevD.74.123507},
  74, 123507

\bibitem[\protect\citeauthoryear{{Tollet} et~al.,}{{Tollet}
  et~al.}{2016}]{2016MNRAS.456.3542T}
{Tollet} E.,  et~al., 2016, \mn@doi [\mnras] {10.1093/mnras/stv2856}, \href
  {https://ui.adsabs.harvard.edu/abs/2016MNRAS.456.3542T} {456, 3542}

\bibitem[\protect\citeauthoryear{{Viel}, {Lesgourgues}, {Haehnelt}, {Matarrese}
   \& {Riotto}}{{Viel} et~al.}{2005}]{2005PhRvD..71f3534V}
{Viel} M.,  {Lesgourgues} J.,  {Haehnelt} M.~G.,  {Matarrese} S.,   {Riotto}
  A.,  2005, \mn@doi [\prd] {10.1103/PhysRevD.71.063534}, \href
  {https://ui.adsabs.harvard.edu/abs/2005PhRvD..71f3534V} {71, 063534}

\bibitem[\protect\citeauthoryear{Wang \& White}{Wang \&
  White}{2007}]{Wang_White_2007}
Wang J.,  White S. D.~M.,  2007, \mn@doi [Mon. Not. Roy. Astron. Soc.]
  {10.1111/j.1365-2966.2007.12053.x}, 380, 93

\bibitem[\protect\citeauthoryear{{Waterval} et~al.,}{{Waterval}
  et~al.}{2022}]{Waterval2022}
{Waterval} S.,  et~al., 2022, \mn@doi [\mnras] {10.1093/mnras/stac1191}, \href
  {https://ui.adsabs.harvard.edu/abs/2022MNRAS.514.5307W} {514, 5307}

\bibitem[\protect\citeauthoryear{{Yang}, {Mo}  \& {van den Bosch}}{{Yang}
  et~al.}{2003}]{2003MNRAS.339.1057Y}
{Yang} X.,  {Mo} H.~J.,   {van den Bosch} F.~C.,  2003, \mn@doi [\mnras]
  {10.1046/j.1365-8711.2003.06254.x}, \href
  {https://ui.adsabs.harvard.edu/abs/2003MNRAS.339.1057Y} {339, 1057}

\bibitem[\protect\citeauthoryear{{Zwicky}}{{Zwicky}}{1937}]{1937ApJ....86..217Z}
{Zwicky} F.,  1937, \mn@doi [\apj] {10.1086/143864}, \href
  {https://ui.adsabs.harvard.edu/abs/1937ApJ....86..217Z} {86, 217}

\makeatother
\end{thebibliography}

%%%%%%%%%%%%%%%%% APPENDICES %%%%%%%%%%%%%%%%%%%%%

\appendix

\section{concentration-mass relation}
\label{subsec: conc_mass}
Instead of writing the NFW density profile given by Eq. \ref{eq:NFW} in terms of $r_{s}$ and $\rho_{s}$, it can be written in terms of the dark matter halo mass and the halo concentration parameters. The halo concentration is defined by:
\begin{equation}
\label{eq:conc}
 c_{\text{vir}} \equiv \frac{r_{\text{vir}}}{r_{s}},
\end{equation}
where $r_{\text{vir}}$ is the virial radius. Eq. \ref{eq:NFW} then becomes:
\begin{equation}
\label{eq:NFW_conc}
 \rho(r) = \frac{\rho_{h}}{3 \left [\ln{(1 +c_{\text{vir}}) - \frac{c_{\text{vir}}}{1 + c_{\text{vir}}}} \right] x \left (\frac{1}{c_{\text{vir}}} + x \right)^2},
\end{equation}
where $x \equiv r/r_{\text{vir}}$ is the fractional distance to the virial radius and $\rho_{h} \equiv \frac{3 M_{\text{vir}}}{4 \pi r_{\text{vir}}^3}$ is the average halo density within the virial radius.

The concentration-mass relation (hereafter c(M)) reflects the assembly history of haloes and is hence a powerful probe of models of structure formation \citep{2020MNRAS.492L..12G}. In standard CDM, the halo concentration has been shown to be a weakly decreasing function of halo mass \citep{dutton2014cold, child2018halo, gilman2020constraints}. From the shape of the power spectrum of matter density perturbations (specifically that the variance increases at low masses), it follows that it is the small-scale perturbations that go nonlinear and undergo gravitational collapse earliest in the Universe, which gives rise to the current standard hierarchical picture of structure formation. The high concentration for low-mass haloes in CDM reflects the fact that those low-mass haloes formed at earlier times when the Universe had a higher background density \citep{navarro1997universal, Maccio2007}. 

We measure the halo concentrations in our CDM and FDM volume simulations of different particle masses, and excluding haloes containing less than 5000 particles. We also verify that, within statistical uncertainties, the expected c(M) relation is obtained for CDM \citep{dutton2014cold}. In contrast to CDM, we find that the FDM c(M) relation is non-monotonic: while FDM halo concentrations follow the expected power-law relation as in CDM for high masses, we observe a drop in the FDM halo concentration for low halo masses. This decline in the halo concentration below a certain mass scale which is a function of the FDM mass is analogous to what has been observed in WDM \citep[e.g][]{Schneider2012,ludlow2016mass}. The expected drop in halo concentration can be directly attributed to the delayed structure formation in FDM (and WDM) compared to CDM. This is because, as discussed before, the small-scale density perturbations below the quantum Jeans scale are stabilized by the quantum potential in FDM. Since the quantum Jeans scale is slowly shrinking with time (refer to Eq.~\ref{eq:Jeans_scale}), those small-scale perturbations do not collapse until a later time in the Universe when the Universe had a lower background density.

\begin{figure}
\includegraphics[width=\columnwidth, trim={0.16cm 0.11cm 0.11cm 0.11cm}, clip] {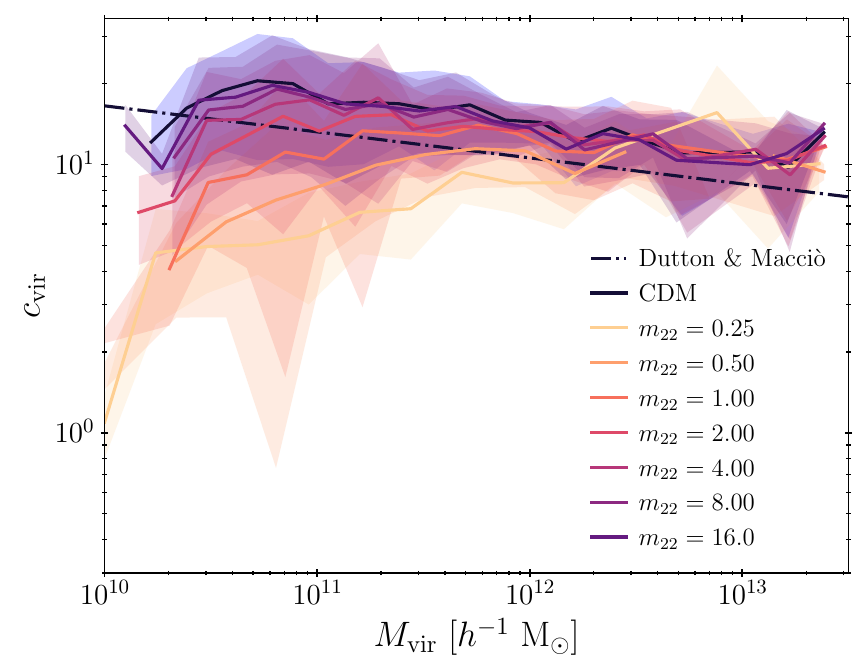}
\caption{The concentration-mass relation for CDM and FDM with different particle masses. The expected power-law relation for CDM is shown in dash-dotted black \citep{dutton2014cold}. The shaded regions show the $1 \sigma$ Poissonian errors for each mass.}
\label{fig:conc_mass}
\end{figure}

%%%%%%%%%%%%%%%%%%%%%%%%%%%%%%%%%%%%%%%%%%%%%%%%%%

% Don't change these lines
%\bsp	% typesetting comment
\label{lastpage}
\end{document}